\documentclass[twocolumn,tighten]{aastex631}
\usepackage{subfigure}

\usepackage{xcolor}

\shorttitle{Spectro-temporal study of Cyg X-2}
\shortauthors{Malu et al.}

\graphicspath{{./}{figures/}}

\begin{document}

\title{A spectro-temporal view of normal branch oscillations in Cygnus X-2 as seen by NICER and NuSTAR}

\author[0000-0003-0440-7978]{Malu Sudha}
\affiliation{Department of Physics \& Astronomy, Wayne State University, 666 West Hancock Street, Detroit, MI 48201, USA}
\author[0000-0002-8961-939X]{Renee M.\ Ludlam}
\affiliation{Department of Physics \& Astronomy, Wayne State University, 666 West Hancock Street, Detroit, MI 48201, USA}
\author[0000-0002-3422-0074]{Diego Altamirano}
\affiliation{School of Physics and Astronomy, University of Southampton, Southampton, Hampshire SO17 1BJ, UK}
\author[0000-0002-8294-9281]{Edward M.\ Cackett}
\affiliation{Department of Physics \& Astronomy, Wayne State University, 666 West Hancock Street, Detroit, MI 48201, USA}
\author[0000-0002-8548-482X]{Jeremy Hare}
\affiliation{NASA Goddard Space Flight Center, Greenbelt, Maryland, 20771, USA }

\affiliation{Center for Research and Exploration in Space Science and Technology, NASA/GSFC, Greenbelt, Maryland 20771, USA }

\affiliation{The Catholic University of America, 620 Michigan Ave., N.E. Washington, DC 20064, USA}

\begin{abstract}
We report the spectro-temporal study of the neutron star low mass X-ray binary Cygnus X-2 using NICER and NuSTAR data while the source was in the normal branch (NB). We detect a normal branch oscillation (NBO) feature at $\sim$ 5.41 Hz that  appears in the middle portion of the NB branch. We note that the NBO appeared only in the 0.5--3 keV energy range, with maximum strength in the 1--2 keV energy band, but was absent in the  3--10 keV energy band of NuSTAR and NICER data. The energy spectrum of the source exhibits an emission feature at $\sim$ 1 keV, previously identified as the Fe L transition in the outer region of the accretion disk. Upon considering both the Fe L and NBO features, we  suggest that the originating location of the Fe L line and the NBOs may coincide and perhaps be due to the same underlying mechanism. Therefore, lags seen in the frequency/energy dependent lag spectra of Cygnus X-2 could be considered to be arising from a region of photoionized material far from the central source. We study the frequency and energy dependent lag spectra of the source, which exhibited a few milliseconds hard lag at the NBO frequency (12-15 ms) and a switch from hard to soft lags at 1 keV. The rms spectrum peaks at 1 keV and the covariance spectrum clearly resembles a thermal spectrum. We discuss the spectro-temporal behavior of the NBO and attempt to constrain its location of origin.

\end{abstract}

\keywords{accretion, accretion disk---binaries: close---stars: individual (Cygnus X-2)---X-rays: binaries}

\section{Introduction} \label{sec:intro}

Neutron Star (NS) low-mass X-Ray binaries (LMXBs) are systems that accrete matter from a low mass ($<$ 1 M$_{\odot}$) companion star, forming an accretion disk around the NS in the process. Based on the characteristic track traced out by these systems in the color-color diagram (CCD) or the hardness intensity diagram (HID) and their timing features, they are broadly classified into `Z' and `atoll' types. Z sources are highly X-ray luminous, with $L_x\sim$ 0.5–1~$L_{\rm Edd}$ and exhibit three main branches in their characteristic Z shape: horizontal (HB), normal (NB), and flaring (FB) branches \citep{hasinger1989}, while atoll are much less luminous with luminosity in the range of $\sim$ 0.001–0.5 $L_{\rm Edd}$ (see \cite{done2007} for a review and references therein) and having three main branches: extreme Island state (hard state), island state (intermediate state), and banana branch (soft state) \citep{hasinger1989,van2006}. The evolution of these sources along their CCD tracks was attributed to $\dot{m}$ variations \citep{hasinger1989,vrtilek1990}. Although other possibilities exist of constant mass accretion rate along the branches with either radiation pressure instabilities at the inner radius or boundary layer (BL) or different accretion flow solutions \citep{homan2002,homan2010,lin2009}. The line between Z and atoll sub classes was blurred when XTE J1701–462 was found to switch between classes at high luminosities, wherein it was concluded that transitions between subclasses are directed primarily by the mass accretion rate component \citep{homan2010}. 

Each of these branches have associated characteristic quasi periodic oscillations or QPOs \citep{hasinger1989}. For Z sources, the HB exhibits QPOs with central frequencies ranging from 15-60 Hz that are referred to as horizontal branch oscillations (HBOs), NB exhibits oscillations with a characteristic frequency of 5-8 Hz \citep{van2006} called NBOs, while FB exhibits 10-25 Hz QPOs called FBOs. The HB and upper NB also exhibits kilohertz QPOs (kHz QPOs; \citealt{van1996}).

There have been several proposed mechanisms for the origin of these QPOs. HBOs could be the beat frequency between the spin frequency of the NS and the Keplerian orbital frequency of the inner accretion disk edge \citep{alpar1985,lamb1989}. Conversely, they could be the nodal precession of tilted orbits near the NS, based on the Relativistic Precession Model (RPM; \citealt{stella1998}). In the RPM, kHz QPOs are considered to be the eccentric orbit's relativistic periastron precession. NBOs associated with the NB of Z sources have more ambiguity associated with their physical process of origin. NBO could be oscillations in the optical depth of a spherical radial inflow due to the radiation pressure feedback of near Eddington luminosity \citep{lamb1989,fortner1989}. \citet{alpar1992} suggest that these are oscillations of sound waves in a thick accretion disk. They have also been proposed to be due to acoustic oscillations associated with a viscous spherical shell surrounding the NS \citep{titarchuk2001}. These models require near Eddington accretion rates, however the NBO-like features observed in atoll sources \citep{wijnands1999,malu2021,mason2023} that are much less luminous than Z sources warrant further investigation of the production mechanism of NBOs. FBOs are considered to share the same production mechanism of NBOs as these oscillations are found to have a smooth transition from NBOs along the NB to FB track \citep{kuulkers1997,casella2006}. The ambiguity and lack of agreement between various models, indicate that a deeper look into the spectro-temporal properties associated with these features is essential. For a more detailed review of QPO theory and interpretation please see \cite{van2006}.

Timing techniques utilizing the Fourier domain are powerful tools to probe the inner accretion disk geometry. Important techniques include cross-spectral analysis \citep{vaughan1997, nowak1999, uttley2014}, cross-correlation analysis and power density spectral analysis. Studying the energy and frequency dependent timing variations (time lags) of the X-ray light can reveal the causal connection between different spectral components and therefore, provide us with information about their physical origin. Energy dependent cross-spectral study allows us to visualize the time/phase lag between photons in two different energy bands for a given frequency range. Most recently, \cite{mendez2024} presented a new technique of simultaneously fitting a combination of Lorentzian functions to the power spectrum and the Real and Imaginary parts of the cross-spectrum that can help measure the lags of weak variability components in NS LMXBs and BH XRBs.

The Fourier cross-spectral analysis of Cygnus X-2 (Cyg X-2) and GX 5-1 was performed by \cite{van1987}. They found hard time lags during HBOs and NBOs along with soft lags in low-frequency noise. Similar studies on Cir X-1 exhibited hard lags (hard photons lagging the soft photons) on the HB and soft lags on the other two branches \citep{kaaret1999}, while the atoll source 4U 1636-536 exhibited soft lags of the order of millisecond timescales \citep{qu2001}. Millisecond hard time lags were explained using the Comptonization process in the corona or jet \citep{vaughan1999,arevalo2006,reig2016}, while the millisecond soft lags were explained using a two layer Comptonization model \citep{nobili2000} or a shot model \citep{alpar1985}. \cite{uttley2011} proposed a disk propagation model to explain the observed lags in black hole (BH) X-ray binaries (XRBs), but modeling the complete spectro-temporal behavior, addressing reverberation delays and propagation delays, was challenging. 
\citet{uttley2023} explain the complex spectro-temporal behavior using a model of the propagation of fluctuations in mass accretion through the disk to an inner corona. A time-dependent Comptonization model is used by \cite{karpouzas2020} to explain variations in the frequency lag spectra, while a modified version of the same model (a variable Comptonization model) is used by \cite{bellavita2022}. Another way to explain the behavior would be to describe it as variations occurring at the outer regions of the accretion flow propagating towards the central object that will modulate the variations from the inner regions, and ultimately modulating the radiation \citep{ingram2013,ingram2011}.

Cross Correlation Function (CCF) studies have shown a few hundred seconds lag in the HB/NB branches of the Z track (e.g., Cyg X-2: \citealt{lei2008}, GX 17+2: \citealt{sriram2019, malu2020}). A few hundred seconds of CCF lags have been associated with the disk readjustment time scales \citep{sriram2012} or more recently, the readjustment timescales of a varying vertical coronal structure \citep{sriram2019}. 

Timing studies can most effectively be used when coupled with the understanding of the energy spectrum of the source. The Comptonized portion in the energy spectrum of Z sources is considered to be due to the hot boundary layer (BL; \citealt{popham2001}) around the NS surface, whereas the optically thick accretion disk is considered to be responsible for the blackbody spectrum. Fourier frequency resolved spectral analysis \citep{gilfanov2003,revnivtsev2006} further supports this model. Generally, the phenomenological prescription used to model the continuum of the energy spectrum of NS LMXBs include a single temperature blackbody and a multicolor disk blackbody for the thermal spectrum, and a powerlaw/broken-powerlaw for the Comptonized hard component \citep{lin2007,cackett2010}. The reprocessed emission from the innermost accretion disk region reveals itself most prominently in the Fe K line region of the energy spectrum. Modeling the reflection spectrum has proven to be a reliable method for providing an upper limit on the NS radius \citep{cackett2008,ludlam2017} along with constraints on other spectral parameters. See \citet{ludlam2024} for a recent review.

Here, we study the spectro-temporal behavior of the NS LMXB Cyg X-2. This highly luminous source, despite being studied extensively over the years, exhibits complex behavior that requires continued investigation. Cyg X-2 is located at $\sim$ 11.3 kpc \citep{ding2021} and is classified as a Z source \citep{hasinger1989}. \citet{mitsuda1989} found time lags ($\sim$ 80 ms) between 1-7 keV and 7-18 keV energy bands, which is basically a 150\mbox{$^{\circ}$} phase lag in the 5.3 Hz QPO. Later, \citet{dieters2000} confirmed this using EXOSAT data, where they note a phase shift in the 5-6 keV energy range, which also corresponded to a minimum in the energy rms spectrum. The NBO and its behavior was explained using the radiation-hydrodynamic model \citep{lamb1989,fortner1989,miller1992}, which states that, at near Eddington accretion rates, the radiation pressure feedback on a spherical radial inflow at the inner disk can induce optical depth fluctuations in the flow, which we observe as the $\sim$ 6 Hz NBOs. The increase or decrease in the flux will depend on the Compton scattering optical depth variations. It was also suggested that if the Compton temperature of the X-ray spectrum is sufficiently low and the Compton scattering optical depth variation in the radial inflow is more than the effect of luminosity changes during the NBO, then we can expect the energy spectrum to show a ``pivot". This would result in the $\sim$ 180$^{\circ}$ phase jump that has been noted (see \citealt{mitsuda1989}). NBO like features have been noted in atoll sources that are much less luminous than the Z sources, which would make it difficult to support the requirement of near Eddington accretion rates for the formation of these oscillations. \citet{miller1992} found that the electron temperature of the oscillating radial inflow is  $\sim$ 1 keV, which is similar to the Compton temperature of the source's X-ray spectrum.

Cyg X-2 also shows excess emission near $\sim$ 1 keV \citep{vrtilek1986,vrtilek1988,kallman1989,smale1993,kuulkers1997} which has been attributed to a Fe L-shell transition. This emission could be due to the photoionization of the accretion disk or possibly accretion disk corona due to the X-ray flux being emitted from the central source/region (e.g. \citealt{schulz2009}). The temperature and density of the illuminated corona and (indirectly the mass accretion rate onto the NS) would be responsible for the strength of this emission feature \citep{liedahl1990,kallman1995}.

There have been only a limited number of studies of the spectro-temporal behavior of Cyg X-2 in the $<$ 1 keV energy domain \citep{jia2023,piraino2002}. The extreme soft energy range from NICER and the simultaneous hard energy information from NuSTAR, along with their high temporal resolution can prove to be very useful in unraveling certain spectro-temporal behavior of such NS LMXBs. In this paper, we discuss the detection and spectro-temporal behavior of an NBO feature in Cyg X-2 using NICER and NuSTAR data.

\section{Observation and Data Reduction}\label{sec:obs}

We use the observations of Cyg X-2 performed by NICER and NuSTAR simultaneously on 2019-09-10, 2019-09-12 and 2022-05-01, which correspond to NICER ObsIDs 2631010101 (exp. $\sim$ 12.7 ks), 2631010201 (exp. $\sim$ 12.1 ks), 5034150103 (exp $\sim$ 4.5 ks) and NuSTAR IDs 80511301002 (exp. $\sim$ 11.3 ks), 80511301004 (exp. $\sim$ 12.7 ks), 30801012002 (exp $\sim$ 15.2 ks) respectively. The data obtained in 2019 was previously analyzed by \citet{ludlam2022}. We follow the standard procedures of data reduction as followed by \citet{ludlam2022}. The NuSTAR Data Analysis Software (NuSTARDAS; \citet{nustardas} ) was utilized for extracting NuSTAR level 1 products. Data calibration and screening was performed using the NuSTAR pipeline `nupipeline' to create the cleaned event files. The level 2 science products were extracted using `nuproducts' task which uses the specified source and background regions. A 100$^{\prime\prime}$ circular region was chosen for source extraction using DS9.
A background region of same radius was chosen which is adequately far from the source for product extraction. Filters used to extract events are the same as those used by \citet{ludlam2022} which takes care of the excess counts due to source brightness. A barycentric correction was performed on all cleaned event files using the `barycorr' tool. The RA/DEC coordinates (RA: 326.17148 Dec: 38.321412) in the FK5 reference frame were used for barycentric correction. Background subtracted energy dependent NuSTAR lightcurves of 20 s time bin were generated for CCF analysis (discussed in Section \ref{sec:results}). NICER data analysis was performed using the `nicerl2' routine following standard procedure. For extracting the spectral and lightcurve products, the `nicerl3-spect' and `nicerl3-lc' routines were used. SCORPEON background models were used for spectral analysis \footnote{https://heasarc.gsfc.nasa.gov/docs/nicer/analysis$\_$threads/scorpeon-overview/}. A barycentric correction was performed on the cleaned event files using the `barycorr' tool. NICER lightcurve background was estimated using the SCORPEON model. Background counts were $<$ 1 count s$^{-1}$, which is negligible compared to the source count rate, hence were not subtracted as per \citet{zhang2021}. NICER lightcurves of 20 s time bin were generated for CCF analysis (discussed in Section \ref{sec:results}).
STINGRAY\footnote{https://docs.stingray.science/en/stable/} \citep{huppenkothen2019,huppenkothen2019a,bachetti2021} was used to obtain all the Fourier products essential for timing analysis, including the cross-spectra and power density spectra in this study.

Following the convention adopted in \citet{ludlam2022}, we henceforth refer to the NuSTAR 80511301002 and NICER 2631010101 observations as obs2, and NuSTAR 80511301004 and NICER 2631010201 observations as obs3. The simultaenous NuSTAR and NICER observarions obtained in 2022 are referred to as obs4. Individual GTIs were selected for segment-wise lightcurve analysis for both NICER and NuSTAR from obs2, obs3 and obs4. For studying the spectral evolution of the source, the normal branch observations from NICER obs2 and obs3 were segmented into 3 epochs (discussed in Section \ref{sec:results}), for which the associated events were extracted using the NICER task - `niextract-events'. Owing to the long gap between the 2019 observations and the 2022 observations, we did not combine these data during the detailed analysis discussed in Section \ref{sec:results} due to long-term overall intensity variations of the source (e.g., fig.\ 1 in \citealt{ludlam2022}). But we did perform an independent energy dependent search for QPOs in this observation (see Section \ref{sec:results}). We note here that a detailed spectro-temporal study like the one that is reported in this work will be performed on other NICER-NuSTAR datasets, for other NS LMXBs in addition to Cyg X-2, as this is part of a large-scale ongoing project on investigating energy dependent temporal variation.

\citet{ludlam2022} performed the reflection spectral modeling of the contemporaneous NuSTAR (80511301002, 80511301004) and NICER (2631010101, 2631010201) observations of Cyg X-2, to constrain the NS radius. Owing to the availability of long duration of overlapping NICER and NuSTAR data, these same observations were chosen for a detailed timing analysis. During obs2 and obs3, Cyg X-2 traced out the HB to NB track of the CCD/HID. The color-color diagram obtained using NICER data (48 s time bin) is as shown in Figure \ref{fig:ccd}, where soft color is the ratio of X-ray photon counts in the 3--5 keV and 2--3 keV energy bands, and hard color is the ratio obtained from the counts in 5--8 keV and 3--5 keV energy bands. For the NuSTAR HID, please refer to Figure 1 in \citet{ludlam2022}. As can be clearly seen, the source is in the lower-intensity state during these observations.

Power density spectra (PDS) were generated for each individual GTI based segments of the lightcurve. Each GTI segment was split into segments of 32s length and for each such segment a PDS was generated with a bin time of 0.0078125 s (i.e., 1/128 s), giving a Nyquist frequency of 64 Hz, and a frequency resolution of 0.03125 Hz (i.e 1/32 Hz). All PDS were then averaged into a final periodogram.  Fractional rms normalization \citep{belloni1990, miyamoto1992} was used to normalize the obtained PDS. The background contribution is negligible ($<$ 1 count/s) for both NICER and NuSTAR, hence was ignored. Theoretical Poisson noise (2/$\mu$; where $\mu$ is the mean count rate) was subtracted from all the obtained NICER PDS. Each PDS was logarithmically rebinned by a factor of 0.02 to lower the noise level.

Complex cross spectra were generated for lightcurve segments of length 32 s each with a bin time of 0.0078125 s, which were then averaged into a final cross spectra following standard techniques \citep{uttley2014} to perform a Fourier frequency resolved time lag study between the different lightcurves. This routine was also used to generate the co-spectra of NuSTAR data using the same bin and segment size as was used for the NICER PDS. NuSTAR observes simultaneously with FPMA and FPMB instruments, providing the ability to produce co-spectra (the real part of cross spectra) for observations. Co-spectra can act as a good estimate of the poisson subtracted PDS, which also takes into account the effect of dead time as well (for a detailed explanation see \citealt{bachetti2015}). Co-spectra of NuSTAR data were used instead of the PDS to search for QPO/QPO-like features. The co-spectra generated was Leahy normalized and logarithmically rebinned by a factor of 0.02. 

\section{Results}\label{sec:results}
\subsection{Timing Analysis}
 
\begin{figure}
\centering
\includegraphics[width=0.49\textwidth, angle=0]{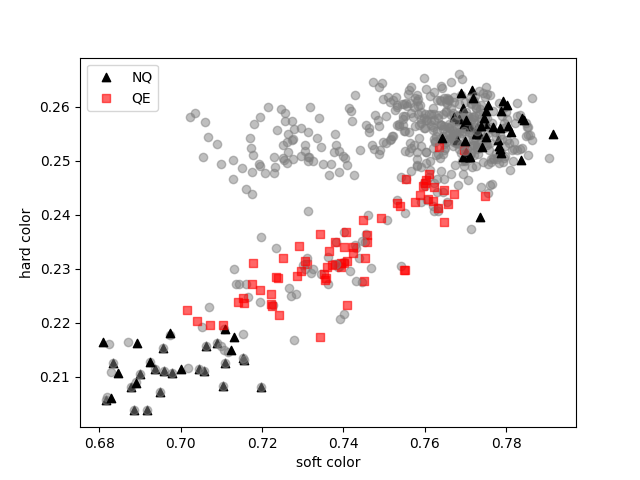}
\caption{ Color-color diagram of Cyg X-2 obtained using NICER obs2 and obs3. Soft color is defined as ratio of X-ray photon counts in the 3--5 keV and 2--3 keV energy bands, and hard color is the ratio obtained from the counts in 5--8 keV and 3--5 keV energy bands. Red squares mark the region that shows signatures of a QPO (QPO epoch; QE)and black triangles mark the region without the presence of a QPO (No QPO epoch; NQ). Grey circles indicate the overall color-color track.}
\label{fig:ccd}
\end{figure} 

NICER and NuSTAR data were analysed segment-wise (each individual GTI) to understand the temporal evolution of the source. The NuSTAR 3--10 keV co-spectra for obs2, obs3 and obs4 showed no significant QPO/QPO-like features with an rms upper limit of $<$ 1.0 \%, 1.1 \%, and 2.3 \% respectively with a 90 \% confidence level.

The corresponding PDS of NICER obs2 in the 0.5-10 keV energy band exhibited signatures of a $\sim$ 5-6 Hz QPO/peaked noise component in a segment (segment 14 in the lightcurve) within a span of $\sim$ 3300 s (duration between the GTI segment that exhibited QPO and the preceding segment which did not show significant signatures of a QPO), which then persisted for the next two consecutive segments and eventually disappeared in a span of $<$ 4000 s (duration between the GTI segment that exhibited a QPO and the following segment which did not show significant signatures of a QPO). The absence of this feature in the simultaneous NuSTAR lightcurve led to an energy dependent search for this feature in the NICER lightcurve. This feature was found to be present only in the 0.5-3 keV energy band, and showed no signatures in the $>$ 3 keV energy range, with an rms upper limit of $<$ 2.2 \% in the NICER 3-10 keV energy band. 

\begin{figure*}
\includegraphics[width=\textwidth, angle=0]{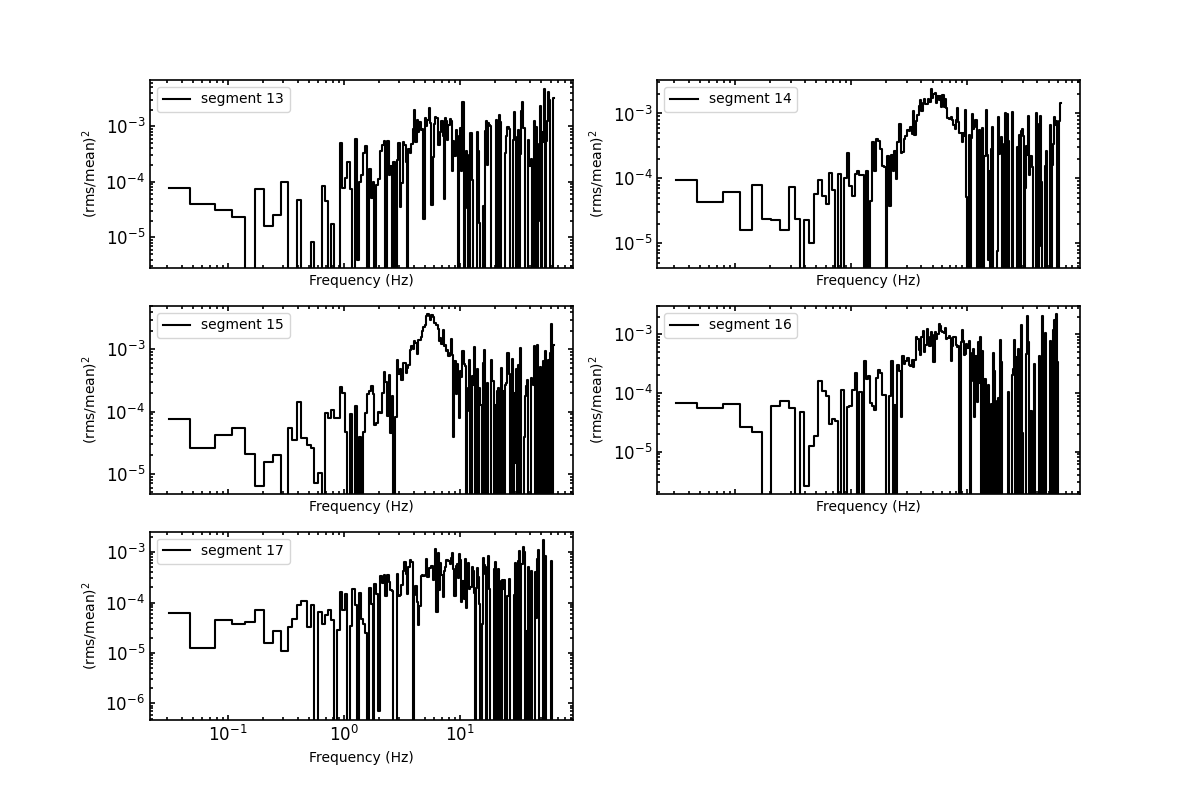}
\caption{NICER PDS of GTI segments in the 0.5--3 keV energy band preceding, during and succeeding the NBO detected in obs2, thus shows the evolution of the NBO in obs2. PDS is Poisson noise subtracted in the figure.}
\label{fig:pds}
\end{figure*}

The gradual appearance and fading out of the feature in the NICER lightcurve of obs2 is shown in Figure \ref{fig:pds}. GTI segments are numbered based on their index from the start of the respective lightcurve. This feature was detected prominently in two segments (segment 14 and 15 as indicated in Figure \ref{fig:pds}).

To verify the presence and significance of the observed QPO in the PDS, a model selection task was employed where we estimate if a noise model (powerlaw + constant) is sufficient to describe the PDS or if it requires an additional Lorentzian term to model any QPO present. The likelihood ratio test (LRT) is used to differentiate between two nested models where the alternate hypothesis is a powerlaw+constant+Lorentzian model and the null hypothesis is a model with the amplitude of the Lorentzian set to zero. Twice the minimum loglikelihood of the two models is quantified as the LRT statistic. 

This LRT statistic can be calibrated via posterior predictive simulations (see \citealt{protassov2002}). For this we used the  {\it emcee} package in python \citep{foreman2013} to run a Monte Carlo simulation in order to generate 1000 periodograms from the simple noise model, which is the null hypothesis. This gives us a sampling distribution of the LRT statistic, which is then used to determine a predictive p-value which is obtained from the tail areas of the simulated LRT distribution. For obs2, we estimate a p-value of 0.42 for segment 13, 0.07 for segment 14, 3.06$\times$10$^{-3}$ for segment 15, 0.41 for segment 16 and 0.56 for segment 17. We consider p-value $\leq$ 0.05 to be strong evidence to reject the null hypothesis and 0.05$\leq$p-value$\leq$0.1 to be a weak indication to reject the null hypothesis. Therefore, segment 14 shows weak/marginal evidence to reject the null hypothesis and segment 15 shows strong evidence for the same. For the other two segments we cannot rule out the null hypothesis. 

To obtain the best fits, PDS of segment 14 and 15 were modeled using a power-law + Lorentzian + constant model, to obtain its central frequency, FWHM and consequently its quality factor Q. The Stingray modeling subpackage was used to obtain the best fit. The LogLikelihood and ParameterEstimation classes in the package were used to obtain the best fit parameters. Modeling via the Loglikelihood class essentially minimizes the log-likelihood function through an optimization algorithm which is equivalent to maximizing the likelihood function (see \citealt{barret2012}).

The central frequency of the QPO in segment 14 was found to be 4.88 $\pm$ 0.50 Hz with an FWHM of 3.34 $\pm$ 1.09 Hz and rms amplitude of 4.45 $\pm$ 1.57 \%. QPO was present in segment 15 at a central frequency of 5.42 $\pm$ 0.47 Hz with an FWHM of 2.35 $\pm$ 1.02 Hz and rms amplitude of 4.69 $\pm$ 2.74 \%. The feature detected in obs2 goes from Q $\sim$ 1.46 to Q $\sim$ 2.31. Values quoted are at 90 \% confidence level.

A similar search for QPOs in the PDS of obs3 led to the detection of a $\sim$ 6 Hz feature in the 0.5-3 keV energy band, which again was absent in the $>$ 3 keV energies in both NICER PDS and NuSTAR co-spectrum. This feature appeared within a span of $\sim$ 3600 s (duration between the GTI segment that exhibited QPO and the preceding segment which did not show significant signatures of a QPO), persisted for the next two consecutive segments, each spanning $\sim$  972 s and $\sim$ 521 s respectively, with a data gap of $\sim$ 419 s between them. Then the feature disappeared in $<$ 3700 s (duration between the GTI segment that exhibited QPO and the following segment which did not show significant signatures of a QPO). The feature is barely visible in segment 3 (3rd GTI in the lightcurve) with a p-value of 0.31 whereas for segment 4 (4th GTI) the p-value was estimated to be 0.004. For the preceding and succeeding segments (i.e., 2nd and 5th GTIs) p-values were found to be 0.57 and 0.67 respectively.

The feature had a central frequency of 5.59 $\pm$ 0.85 Hz, an FWHM of 2.87 $\pm$ 1.52 Hz and rms amplitude of 3.18 $\pm$ 1.70 \% when it was first barely visible in obs3 (segment 3) with a Q factor of $\sim$ 1.94. Following this segment, it was found to have a Q factor of 2.6, with a central frequency of 5.59 $\pm$ 0.77 Hz, an FWHM of $\sim$ 2.59 $\pm$ 0.49 Hz and rms amplitude of 4.08 $\pm$ 2.08 \%.

Similarly, in obs4, out of the 3 GTI segments that has a simultaneous NuSTAR observation, one of them showed signatures of a QPO feature at $\sim$ 6 Hz which was present in the 0.5--3 keV energy band with an rms amplitude of 4.93 $\pm$ 2.28\% and absent in the 3--10 keV energy band. As explained previously, this PDS was fitted to obtain a QPO central frequency of 6.32 $\pm$ 0.83 Hz, FWHM of 1.65 $\pm$ 1.06 and therefore a Q value of $\sim$ 3.83 . The feature was detected with a p-value of 0.03 which gives strong evidence for the rejection of the null hypothesis.

To understand the nature of these detected QPOs, the specific location of the GTIs exhibiting QPO signatures were located in the color-color diagram (Figure \ref{fig:ccd}). It can be clearly noticed from the diagram obtained for obs2 and obs3 that the QPO appears in the middle portion of the Normal Branch (NB), which along with their central frequency and Q factor, suggests that they are Normal Branch Oscillations (NBOs). Segments immediately preceding and following the QPO, were found to be in the upper NB and lower NB, thus clearly correlating with the spectral state evolution of the source. Obs4 showed a single branch in the color-color diagram rendering it difficult to ascertain the spectral state of the source.

\begin{figure}
\centering
\includegraphics[width=0.48\textwidth, angle=0]{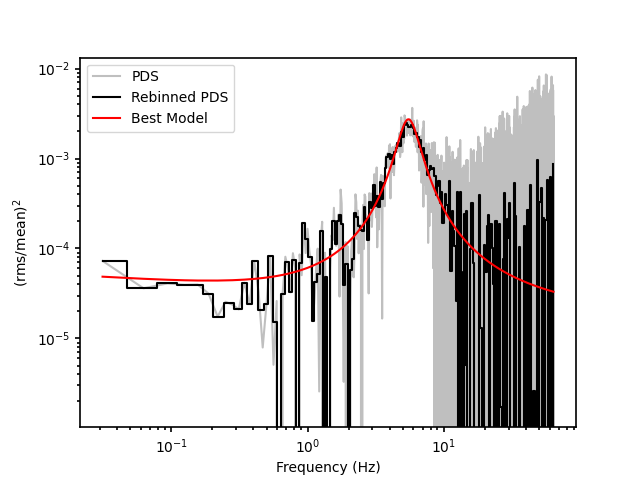}
\caption{Poisson noise subtracted averaged PDS of the NICER  QE epoch in the 0.5--3 keV (grey line) logarithmically rebinned by a factor of 0.02 (black line) modeled using a  power-law + Lorentzian + constant model. Here, the red line indicates the best fit obtained.}
\label{fig:qpofit}
\end{figure}

Since there is no evidence for the QPO central frequency evolving over time between obs2 and obs3 (i.e., central frequency is consistent within errors), segments that exhibited QPOs lying in the middle of the NB in these observations were combined (referred to as QE (QPO epoch) hereafter) and an averaged PDS was obtained (Figure \ref{fig:qpofit}), which was then modeled using a power-law + Lorentzian + constant model. The best-fit resulted in a central frequency of 5.41 $\pm$ 1.02 Hz, FWHM of 2.10 $\pm$ 1.01 Hz and consequently, a quality factor of 2.57 $\pm$ 1.31. Obs4 was not combined with obs2 and obs3 owing to the large gap between the datasets. Only the combined dataset is used for performing further in-depth spectro-temporal analysis.

The rms upper limit at the QPO frequency in the Leahy normalized NuSTAR co-spectra and NICER 3--10 keV PDS was estimated by fitting the co-spectra/PDS with the same model after fixing the Lorentzian central frequency and Lorentzian width to the best fit values from the NICER 0.5--3 keV PDS of the QE epoch, but by allowing the Lorentzian normalization to vary. The 90\% confidence upper limit of the Lorentzian normalization parameter was then used to determine the rms upper limit. The equation $\sqrt{\frac{\pi  \cdot LN \cdot LW}{2 \cdot \text{(mean countrate)}}} $ was used (e.g., \citealt{pasham2013,lazar2021}) to determine the rms upper limit, where LN and LW stand for Lorentzian normalization and width.

The rms upper limit at the QPO frequency in the NICER 3--10 keV energy band (QE) was estimated to be $<$ 2.2 \% with a 90\% confidence level. For obs4 this value was found to be $<$ 1\%. The rms upper limit at the QPO frequency in the NuSTAR 3--10 keV energy band data was estimated to be $<$ 1.0\% (obs2), 1.1\% (obs3) and $<$ 2.3\% (obs4) with a 90\% confidence level. 

Furthermore, to determine the sensitivity of QPO detection we used the equation (e.g.  \citealt{belloni2012}):
\begin{equation}
n\sigma = \frac{1}{2}\left(\frac{S^2}{B+S}\right)r_s^2 \left(\frac{T}{\Delta \nu}\right)^{\frac{1}{2}}
\end{equation}
where  where S, B, r$_s$ are the source count rate, background count rate and fractional rms variability respectively. T is the exposure time and $\Delta$$\nu$ the FWHM of the QPO. We find that in the 3-10 keV energy band a QPO of 1 \% rms amplitude could be detected with $>$ 1$\sigma$ significance and a 2 \% rms amplitude QPO could be detected with $>$ 3$\sigma$ significance. Since a 2.2\% rms upper limit was estimated in the 3-10 keV energy band of NICER, we can conclude that a QPO of the given rms variability and $\Delta$$\nu$ obtained from fit would have been detected if it were present.

The lightcurve segments preceding and following the epochs/segments were QPOs are seen are named as NQ1 (lying in the upper NB) and NQ2 (lying in the lower NB). NQ stands for ``No QPO" in this context. These NQ1 and NQ2 epochs were independently selected and combined to create a representation of the upper NB and lower NB to perform further comparative spectral studies. Further timing analysis was performed for the QE epoch lying solely in the middle of the NB.

\begin{figure*}
\includegraphics[height=0.30\textwidth,width=7.8in, angle=0, trim = 170 0 0 0, clip]{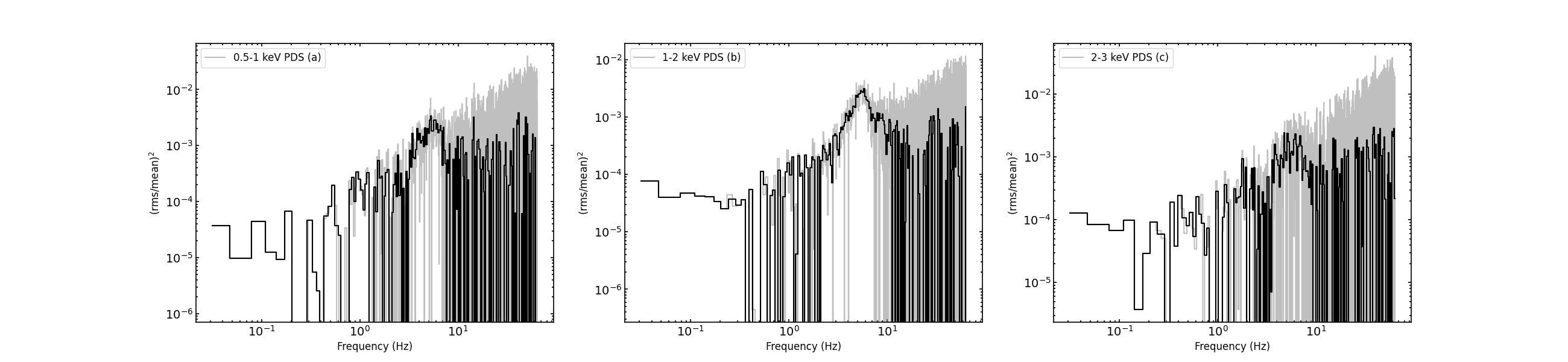}
\caption{Panels (a), (b) and (c) show PDS of the NBO segment (QE) in 0.5--1 keV, 1--2 keV and 2--3 keV energy bands respectively. The grey line indicates the obtained averaged PDS and the black line indicates the averaged PDS logarithmically rebinned by a factor of 0.02. It can be clearly noted that QPO peaks in strength in the 1--2 keV energy band. All the PDS shown are Poisson noise subtracted.}
\label{fig:rms}
\end{figure*}

A more detailed search for this QPO feature was performed within the 0.5--3 keV energy band by further dividing this energy band into 0.5--1 keV, 1--2 keV and 2--3 keV energy bins. The QPO peaks in the 1--2 keV energy band with a p-value of 0.04 as can be seen in Figure \ref{fig:rms}. In the 0.5--1 keV and 2--3 keV energy bands a p-value of 0.1 was noted which indicates a weak feature. Implications of this finding is discussed in Section \ref{sec:discussion}. 

Cross-spectral timing analysis was performed on the QE epoch exhibiting NBO (for a recent review of spectro-temporal techniques, see \citealt{uttley2014}). Energy bands used for the analysis were chosen such that they represent the energy range where the NBO is detected.

\begin{figure}
\centering
\includegraphics[width=0.48\textwidth, angle=0]{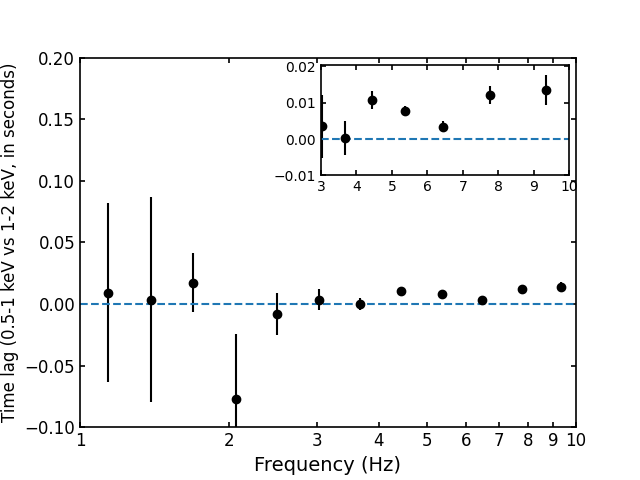}
    \caption{Frequency lag spectra obtained in the 0.5--1 keV vs 1--2 keV energy bands. Inset shows the 3-10 Hz frequency range of the spectra to highlight the lag occurring in the NBO frequency range. Dashed line indicates 0 lag, plotted to clearly indicate the soft and hard lags. Soft lags indicate that soft photons are lagging the hard photons (negative lag values here) and vice versa for hard lags.}
    \label{fig:freqlag}
\end{figure}

Figure \ref{fig:freqlag} shows the frequency dependent lag spectra highlighting the frequency range where the NBO is present. In the frequency-lag spectra for the energy bands 0.5--1 keV vs. 1--2 keV, we notice near zero lags in the 1--2 Hz frequency range. In the 2--3 Hz frequency range, we see a trend towards soft lags (negative lags) which are then seen to be evolving again towards near zero lags indicated by the dashed line in Figure \ref{fig:freqlag}. In the 4--9 Hz frequency range, where we detect the NBO, we notice 12--15 ms hard lags shown in the inset. Implications of this result are discussed in Section \ref{sec:discussion}.

\begin{figure}
\centering
\includegraphics[width=0.48\textwidth, angle=0]{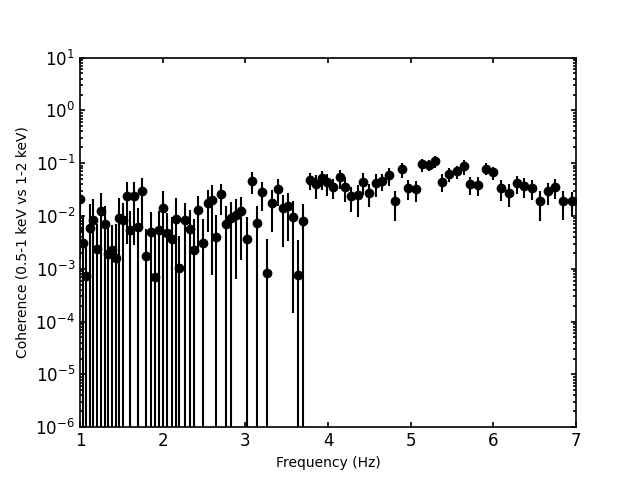}
    \caption{Coherence as a function of Fourier frequency obtained in the 0.5--1 keV vs 1--2 keV energy bands.}
\label{fig:coherence}
\end{figure}

To investigate the degree of correlation between variations in the 0.5--1 keV vs 1--2 keV lightcurves, we estimated the coherence function \citep{vaughan1997}. Coherence was determined as a function of frequency (Figure \ref{fig:coherence}). Centered around the QPO central frequency where we note a lag of around 12--15 ms, the coherence value is $\sim$ 0.04, which is above the statistical threshold coherence estimate of 1.2/(1+0.2m) \citep{epitropakis2017} where m is the number of lightcurve segments (here m=210). It can be seen in the plot that coherence is significantly better constrained over the range of NBO $\nu$$_{qpo}$ $\pm$  FWHM/2 although the coherence estimate is low (similar to Figure 5 in \citealt{mitsuda1989}). The loss of coherence could possibly be attributed to the low quality factor of the NBO.

To probe the energy dependent behavior of the NBO, lags centered around the centroid frequency of the NBO were studied, such that the frequency interval spans within $\nu$$_{qpo}$ $\pm$ FWHM/2 (e.g., \citealt{wijnands2001,cackett2016,troyer2018,jia2023}). 

\begin{figure}
\centering
\includegraphics[width=0.48\textwidth, angle=0]{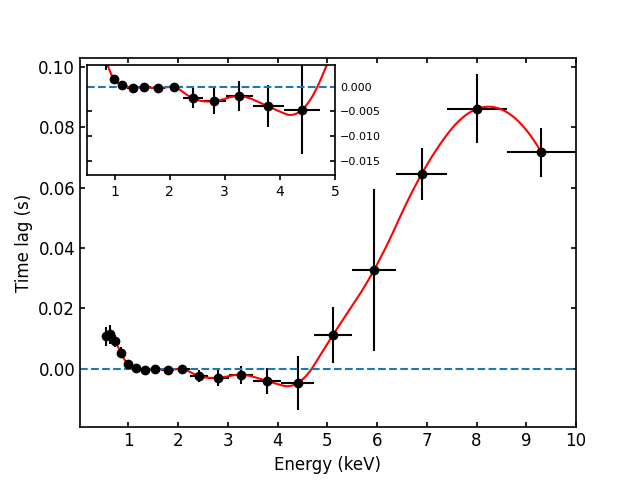}
    \caption{Energy lag spectra obtained in the frequency interval centered around the NBO central frequency. Dashed line indicates 0 lag. Inset shows the lag obtained in 0.5--3 keV energy range where the NBO is present. A sign reversal (hard to soft) is seen in the lags at around 1.3 keV and a sign reversal is seen again at around 5 keV.}
    \label{fig:energylag}
\end{figure}

Lags were extracted between 20 small energy bins spanning the 0.5--10 keV energy band and by taking 1--2 keV as the reference (Figure \ref{fig:energylag}). The frequency interval was centered at the NBO peak frequency with a range across $\nu$ $\pm$ FWHM/2. The lag energy spectra for the QE epoch is as shown in Figure \ref{fig:energylag}. Below 1.2 keV there is a clear detection of a hard lag. At $\sim$ 1.2 keV we see that the lags initially evolve towards zero and then shows a trend towards negative lag values till around 4.5 keV which can not be confirmed owing to the uncertainties on these values. Above 4.5 keV there is a sign reversal and the lags evolve to the hard regime. To check on the dependence of the trend seen in the lag evolution on the energy interval chosen as reference band, we used 0.5--3 keV energy band as the reference and recreated the energy dependent lag spectra. We note that the overall trend are consistent within uncertainties. Our results are in close agreement with the results from \citet{jia2023} where they see a hard lag below 1.5 keV and also in agreement with the studies of \citet{dieters2000} and \citet{mitsuda1989} where they see a phase jump in the 5--6 keV energy range. The inset in Figure \ref{fig:energylag} shows the segment of the lag spectra where the NBO is detected (0.5--3 keV energy range). Section \ref{sec:discussion} discusses the interpretations of this finding, by coupling these results with the energy spectral analysis results that will be discussed in Section \ref{sec:results}.

\begin{figure}
\centering
\includegraphics[width=0.48\textwidth, angle=0]{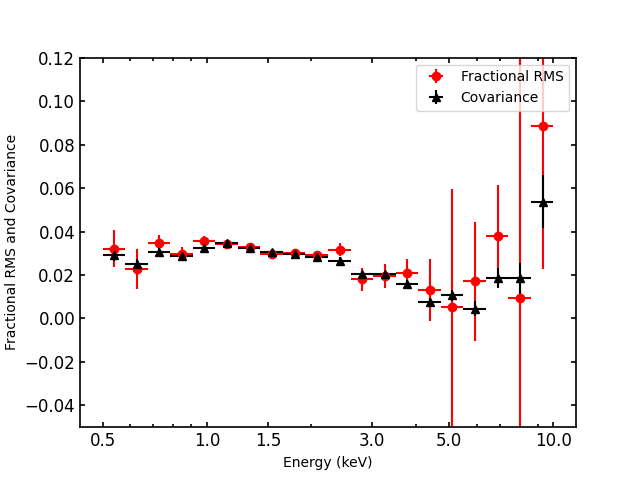}
    \caption{Fractional RMS and covariance spectrum obtained in the 0.5--10 keV energy band between a frequency interval centered around the NBO central frequency.}
    \label{fig:rmscov}
\end{figure}

In an attempt to further understand the physical origin of the NBO, a QPO rms spectrum was generated using the RmsSpectrum object in Stingray (Figure \ref{fig:rmscov}). The rms spectrum was obtained in the 0.5--10 keV energy band between a frequency interval centered around the NBO central frequency as mentioned above. The QPO spectrum peaks in the $\sim$ 1--1.5 keV in the 0.5--3~keV energy band. 
As previously suggested, we do see a tentative minimum in the fractional rms around the pivot energy but the uncertainties on the values prevent a confirmation of the same. The large uncertainties above 5 keV is due to the effective area of NICER declining steeply at higher energies. A covariance spectrum (see \citealt{uttley2014}) was also generated to better understand and isolate the spectral components contributing to the variation (see Figures \ref{fig:rmscov}).
The covariance spectrum essentially provides a description of the spectral shape of energy bins of interest,  which is correlated with the reference band. It can be considered as equivalent to the rms spectrum. The CovarianceSpectrum class in Stingray was used to obtain the covariance spectra in the 0.5--10 keV energy band, averaging over the frequency range centered around the NBO central frequency. The obtained covariance spectrum was then folded with the NICER response to model it similar to the energy spectra using XSPEC v12.13.1 \citep{arnaud1996}. We find that the covariance spectrum resembles a thermal spectrum, which can be described using a single temperature blackbody ($\sim$ 2.8 keV) and a multicolor blackbody ($\sim$ 0.70 keV) model. A discussion on the implications of this result can be found in Section \ref{sec:discussion}.
\\
\subsection{Spectral analysis}   \label{sec:spect}
A detailed NICER+NuSTAR spectral analysis of this dataset using two independent reflection models has been previously performed and reported by \cite{ludlam2022}. In this study, we analysed the NICER+NuSTAR spectra (0.5--20 keV) of the three epochs (NQ1, QE and NQ2) using XSPEC v12.13.1 \citep{arnaud1996}. Also, since the intention of this analysis is to understand spectral variations before, during and after the occurrence of the NBO, emphasis was laid upon the phenomenological continuum modeling of the spectra and we refrain from an attempt to perform reflection modeling, and refer to the previous results from \cite{ludlam2022}. 

We use {\sc crabcor*tbfeo(diskbb+bbody+powerlaw)} to model the continuum, which resulted in residuals around $\sim$ 1.1 keV and 6.7 keV, indicating emission features of Fe L and Fe K lines as was noted in \cite{ludlam2022}. Following the previous analysis of the dataset, we modeled the $\sim$ 1 keV residual in the NICER spectra with the mekal model, which is a collisionally ionized plasma model \citep{mewe1985,mewe1986,liedahl1995}. The material density is fixed at 10$^{15}$ cm$^{-3}$ \citep{ludlam2022,schulz2009}. A Gaussian model was used to account for the Fe K line. The centroid energy of the Gaussian was fixed at 6.7 keV. The oxygen abundance in the tbfeo model was allowed to vary to obtain the best fit. We fixed the iron abundance to solar abundance to obtain the best fit (abundances were set to wilm values; \citealt{wilm2000}).

Best fit model resulted in $\chi^{2}$/dof values of 306.38/304 in the NQ1 epoch (upper NB), 208.54/297 in the QE epoch (middle NB) and 268.11/298 in the NQ2 epoch (lower NB) (see Table \ref{tab2}). This fit suggests that the Disk temperature $kT_{in}$ has decreased from 1.74 $\pm$ 0.03 keV in the upper NB to 1.52 $\pm$ 0.02 in the lower NB. {\sc diskbb} normalization has increased from  $\sim$ 107 in the upper NB to $\sim$ 163 in the lower NB. Although we see a decrease in  $kT_{in}$ and {\sc diskbb} normalization between NQ1 and QE epochs (upper to middle NB), these values are consistent between the QE and NQ2 epochs (middle and lower NB).
It can be noted that the blackbody temperature has varied from upper to middle NB (NQ1 to QE) but is consistent within error bars between upper and lower NB. Blackbody normalization has significantly varied from middle to lower NB, while it remains consistent within error bars from upper to middle NB. Based on our model, disk and blackbody components are found to be the only component parameters that are varying along the NB. {\sc diskbb} normalization is proportional to the inner disk radius and hence it could be said that the inner disk radius is increasing as we go down the NB, but since this parameter also depends on the color correction factor, we refrain from making any assumptions about the disk radius. 

A higher blackbody temperature just before the appearance of the QPO could also imply a higher degree of radiation pressure from the source which causes radial oscillations in the accretion flow, thus leading to the appearance of the QPO. After the appearance of the QPO the reduction in blackbody normalization could possibly be indicative of a smaller BL and lesser radiation pressure feedback, thus explaining the disappearance of the QPO. But we cannot rule out that the variations in disk and blackbody parameters are caused by degeneracy between the models. Hence, we would refrain from making definitively conclusive statements about the physical implications of such variations.

\begin{figure}
\centering
  
\includegraphics[width=0.5\textwidth, angle=0]{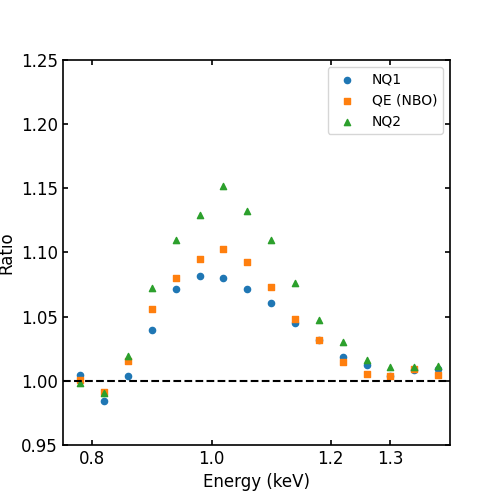}
    \caption{Ratio of NICER data to continuum model indicating the presence of a Fe L line emission at $\sim$ 1.1 keV. The continuum is modeled by a {\sc diskbb+bbody+powerlaw} model. Variation in the Fe L strength as it moves along the NB can be clearly noted.}
    \label{fig:iron}
    \end{figure}

\section{Discussion} \label{sec:discussion}

Our broadband spectro-temporal study of Cyg X-2 in the NB using NICER and NuSTAR led to the detection of a NBO feature at $\sim$ 5.41 Hz in the middle portion of the NB. The feature was found to be present only in the 0.5--3 keV energy band, where it peaked in the 1--2 keV energy range. Frequency and energy dependent lag study of the feature showed a few millisecond hard lag and a switch from hard to soft lags at $\sim$ 1 keV. The energy spectrum of the source shows an excess emission near 1 keV, identified as the Fe L transition line which was modeled using a mekal model with kTe $\sim$ 1.1 keV. Spectral modeling showed variations primarily in the soft disk component parameters, but we note that the powerlaw normalization has varied  with its index remaining consistent within error bars along the NB. A discussion on the spectro-temporal behavior of the source based on these results can be found in the following sections.

 \subsection{NBOs: Spectral energy dependence and the Fe L line}

The only previous study on spectro-timing properties of Cyg X-2 extending down to 0.5 keV was reported by \cite{jia2023} using a different epoch of NICER data in the 0.4--10 keV energy band when the source was in its NB and FB branches of the HID. Their study showed the presence of an NBO feature at $\sim$ 5.8 Hz in a few orbits in the 0.4--10 keV energy band, similar to this study. 

\begin{table*}
\caption{Best-fit spectral parameters for the upper (NQ1), middle (QE) and lower (NQ2) portions of the NB from 0.5--20 keV of NICER+NuSTAR spectra, fitted using  {\sc Crabcor*Tbfeo(diskbb + bbody + Gaussian(Fe) + mekal + Powerlaw)} model. Energy of Fe line here is fixed at 6.7 keV.
The subscript BB represents the {\sc bbody} model and dBB represents {\sc diskbb} model. The flux (absorbed) reported is in the units of 10$^{-8}$ ergs cm$^{-2}$ s$^{-1}$ and is calculated in the 0.5--20 keV energy band. Errors are quoted at a 90\% confidence level. Mekal density was fixed at 10$^{15}$ cm$^{-3}$, mekal abundance at 2 and Tbfeo iron abundance was fixed at solar abundance \citep{ludlam2022}.}
\label{tab2}
\centering
\scriptsize
\begin{tabular}{ccccccccc}
\hline
\hline
\hline
Parameters&NQ1&QE&NQ2&\\
\hline
C$_{FPMB}$ &0.98 $\pm$ 0.02 & 0.96 $\pm$ 0.01& 0.98 $\pm$ 0.01\\ \\
C$_{NICER}$ &0.83 $\pm$ 0.02& 0.79 $\pm$ 0.02 & 0.81 $\pm$ 0.02 \\ \\
$\Delta_{\Gamma}$ &-0.06 $\pm$ 0.01 & -0.11 $\pm$ 0.01& -0.11 $\pm$ 0.01\\ \\
$N_{H}$ ($\times$10$^{22}$ cm$^{-2}$)&0.45 $\pm$ 0.01 & 0.44 $\pm$ 0.01 & 0.46 $\pm$ 0.01\\ \\
Tbfeo A$_{O}$ & 0.65 $\pm$ 0.01 &0.67 $\pm$ 0.01 & 0.67 $\pm$ 0.01\\ \\
$kT_{in}$ (keV)\footnote{Temperature of the {\sc diskbb} model.} &1.74 $\pm$ 0.03& 1.56 $\pm$ 0.02 & 1.52 $\pm$ 0.02 \\ \\
$N_{dBB}$\footnote{Normalization of the {\sc diskbb} model.}& 107.67 $\pm$ 6.78 & 162.07 $\pm$ 7.75 & 163.20 $\pm$ 6.90 \\ \\
$kT_{BB}$ (keV)\footnote{Temperature of the {\sc bbody} model.} & 2.74 $\pm$ 0.09 &  2.46 $\pm$ 0.06 &  2.66 $\pm$ 0.12 \\ \\
$N_{BB}$ ($\times10^{-2}$) \footnote{Normalization of the {\sc bbody} model}& 4.37 $\pm$ 0.35  & 3.88 $\pm$ 0.20 & 1.49 $\pm$ 0.10\\ \\
$kT_{mekal}$ (keV)\footnote{Plasma Temperature of the mekal model.} &1.09 $\pm$ 0.01 &1.11 $\pm$ 0.01 & 1.13 $\pm$ 0.07 \\ \\
$N_{mekal}$\footnote{Normalization of the mekal model.}& 0.11  $\pm$ 0.01 & 0.15 $\pm$ 0.01 & 0.23 $\pm$ 0.03\\ \\
$E_{Fe}$ (keV) \footnote{Line Energy of the Gaussian model for Iron line.}&6.7 &6.7 &6.7 \\ \\
$\sigma_{Fe}$\footnote{Line width of the Gaussian model for Iron line.}& 0.24 $\pm$ 0.20 & 0.34 $\pm$ 0.07 & 0.28 $\pm$ 0.06\\ \\
$N_{Fe}$ ($\times10^{-3}$)\footnote{Normalization of the Gaussian model for Iron line.}& 4.79 $\pm$ 2.90 & 5.52 $\pm$ 1.50 & 6.01 $\pm$ 1.00  \\ \\
$\Gamma_{pl}$\footnote{Power-law index.}& 3.61 $\pm$ 0.07 & 3.66 $\pm$ 0.06 & 3.77 $\pm$ 0.06\\ \\
$N_{pl}$\footnote{Normalization of the {\sc powerlaw} model.}& 2.45 $\pm$ 0.11 & 2.76 $\pm$ 0.11&3.42 $\pm$ 0.11\\ \\
Flux (0.5--20 keV) \footnote{Absorbed flux }& 2.07 $\pm$ 0.01 & 1.98 $\pm$ 0.01& 1.73 $\pm$ 0.01\\ \\
$\chi^{2}$/dof & 306.38/304 & 208.54/297 & 268.11/298\\ \\

\hline
\end{tabular}
\end{table*}

We note from our simultaneous NuSTAR and NICER study that the NBO oscillation was present only in the 0.5--3 keV energy range and the PDS above 3 keV shows no indications of a QPO component with an rms upper limit of $<$ 2.2\%. This was confirmed using both the NICER 3--10 keV and NuSTAR energy band, where rms upper limit at the QPO frequency in the NuSTAR 3--10 keV data was found to be $<$ 1.1 \%. A more specific search for the NBO within the 0.5--3 keV energy band led to a finding that the NBO peaks in the 1--2 keV energy band with a $\sim$4.56\% rms amplitude compared to a $\sim$3.83\% rms amplitude in 2--3 keV energy band.

Interestingly, the energy spectra for the source in 1--2 keV energy range is dominated by an excess emission modeled using the mekal model and this emission is considered to be the signature of an Fe L emission line at $\sim$ 1 keV (see figure 3 from \citet{ludlam2022} and Figure \ref{fig:iron} in this paper). The Fe L line could possibly be a blend of features from the highly ionized states of Fe, Ne, and O that is considered to be emanating from a low density, collisional plasma, where the material of origin is photoionized rather than collisionally ionized, having a temperature $>$ 5 $\times$ 10$^{6}$ K and located far from the inner region of the accretion disk \citep{vrtilek1986}. This could happen when the radiation from the central region photoionizes the cold gas which should be further out in the disk (e.g. \citealt{degenaar2013}). In the QE epoch, the best fit resulted in a plasma temperature of $\sim$  1.1 keV. 

As mentioned in Section \ref{sec:intro}, one of the possible explanations for 6 Hz NBOs comes from the radiation hydrodynamic model \citep{lamb1989,fortner1989,miller1992}. \cite{fortner1989} suggests that NBOs arise due to the radial oscillations in a spherical accretion flow due to the radiation pressure feedback from the central source.  Theoretically, the oscillations at a radius of $\sim$ 300 km from the NS is expected to produce the $\sim$ 6 Hz NBO \citep{fortner1989}.

Given that the $\sim$1 keV emission feature originates further out in the disk and the aforementioned origin of the NBO, we postulate that they may be caused by the same underlying mechanism in this case. In other words, the NBOs are in fact oscillations occurring at a few hundred km away from the central source in a photoionized plasma material.  We also note that the NICER spectra in the $<$ 3 keV band could be modeled satisfactorily with a {\sc diskbb}, blackbody and mekal model. 
Although we discuss an alternative scenario for the NBO in Section \ref{nboalt}.

We must note that NBOs have also been observed in Z sources in the higher energy bands ($>$ 3 keV; e.g. \citealt{mitsuda1989,dieters2000,malu2020,malu2021}). This, when coupled with studies that have found NBO like oscillations in Atoll sources \citep{wijnands1999,reig2004} that have much lower overall luminosity, suggest that the NBO strength possibly depends on the properties of the inflow and Compton scattering cloud.

Moving down along the NB, the Fe L emission appears to become stronger (Figure \ref{fig:iron}). Studies suggest that the optical depth of the radial flow region increases as the source moved down the NB. The optical depth variation could be leading to ionization changes that in effect results in the Fe L emission variation. These same optical depth variations could also be the origin mechanism of NBOs as discussed above. But we must also note that the normalization of the mekal model is maximum in the NQ2 epoch where QPO was not detected which appears counter-intuitive to the scenario explained above. But we must note that since the normalization of the Fe L line estimated using the mekal model also depends on the angular diameter distance to the plasma region of origin, we could assume that a slight increase in the distance could be leading to an increase in the normalization value. If that is the case, then it would mean the location of origin of this Fe L feature has expanded/changed making it no longer coincide with the location of origin of the NBO. There could also be other effects in play here such as the presence of collisional ionization in the plasma in combination with the photoionization of the medium. We must emphasize that we can not make a conclusive statement based on just the current results.

\begin{figure}
\centering
\includegraphics[width=0.48\textwidth, angle=0]{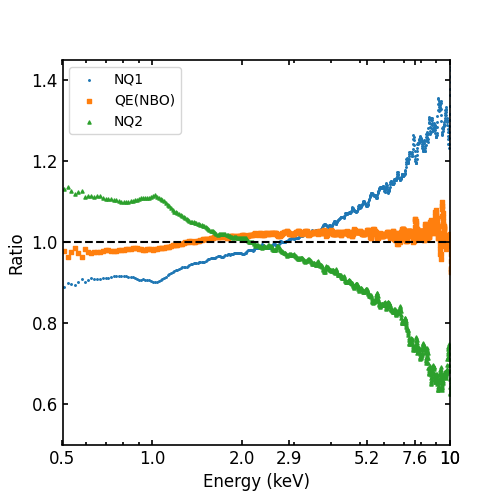}
\caption{Ratios of the unfolded NICER spectra of the NQ1, QE and NQ2 epochs with respect to the total spectrum which exhibits spectral deviations of the QE epoch at 1.5--1.8 keV and $\sim$ 3 keV.}
\label{fig:specratio}
\end{figure}

Upon plotting the ratios of the unfolded NICER spectra of the NQ1, QE and NQ2 epochs with respect to the  averaged spectrum, a deviation from the expected spectral trend at $\sim$ 1.5--1.8 keV is noted in the QE epoch following which it deviated once again at $\sim$ 3 keV (Figure \ref{fig:specratio}). This could mean that the nature of the NBO could be changing at the energies at which the spectrum deviates from the expected trend in the energy spectrum. 
The obtained results cannot tightly constrain these equivalencies, hence future studies of this aspect of the spectro-temporal variation would be of interest.

Furthermore, it has been suggested that NBO frequency increases towards the FB to $\sim$10--20 Hz, becoming a FBO \citep{wang2016}. This possibly means the same origin mechanism for NB and FB. Based on the above consideration of a varying optical depth along the NB, we speculate that NBOs and FBOs could be the same oscillation arising from different radial zones or locations. Oscillations in the spherical inflow could be radially varying with respect to its location, possibly moving to higher frequencies as they move inwards. Since we do not have data from the FB in the datasets selected for this study, we can not investigate this further in this work. It could be further investigated in the FB branch in other datasets, which is outside the scope of this work.

\subsection{NBOs: A time lag study}

In the frequency dependent lag study we note a hard lag $\sim$ 12--15 ms in the 4--9 Hz frequency range where we detect the NBO. A hard lag indicates that the hard photons are lagging behind the soft photons. Upon studying time lags in QPOs, \citet{stollman1987} suggested that Compton scattering through a cloud of oscillating optical depth, could result in oscillating emergent X-rays, where the signal in the higher energy band lags behind that in the lower energy band. This could be explained based on the fact that higher energy photons undergo more scattering, at any given instant, than the lower energy photons. They also suggest that the amplitude of the signal will be larger in the higher energy band, because photons can reach higher energies only if the optical depth is large enough. 

We obtained the energy lag spectra in the 0.5--10 keV energy band using 1--2 keV as the reference energy band (Figure \ref{fig:energylag}). The lag spectrum was calculated over the $\nu$$_{qpo}$ $\pm$ FWHM/2 frequency range. The lag spectrum shows a hard lag below a critical turning point of $\sim$ 1.1--1.2 keV, after which it evolves towards zero lags till 2 keV and then showing an indication of a reversal of sign (soft lags) which can not be claimed owing to the uncertainties associated with these values. This behavior is reflected in the rms-energy spectra where the rms is found to increase until $\sim$ 1 keV and then decrease towards higher energies until 3 keV (Figure \ref{fig:rmscov}). The expected behavior of a phase jump in the lag spectrum is seen in this study at $\sim$ 4.5 keV as has been noted in previous studies \citep{mitsuda1989,jia2023}. \citet{jia2023} use a dual corona Comptonization model to explain the behavior below and above 1.5 keV where they found the roll over. They suggest that there is a phase shift between two Comptonized components that they use to model below and above the turning point, which leads to the roll over. Based on our findings in the 0.5--3 keV energy band where the NBO is present, we suggest that the zero lag in the 1-2 keV energy band could be because of stronger NBO presence in this energy band. A stronger NBO (higher rms) indicates a more localized stable oscillation in the respective energy band, suppressing any lags. The behavior above 3 keV following the expected trend could be associated with a variation of the Comptonization region.

\subsection{NBO: An alternate scenario} \label{nboalt}

Another proposed model for NBOs suggest that these could be the viscous oscillations of a spherical shell surrounding the NS surface within 1 NS radius from the central source \citep{titarchuk2001}, indicating that it could be the BL. In general, the blackbody model is used to account for the hot boundary layer and usually has a temperature $\sim$ 2--3 keV \citep{gilfanov2003}. Variations seen in the blackbody parameters could be indicative of a varying BL, which aligns with this alternate scenario. But the disk parameters shows variation too and therefore, the disk and blackbody parameter variations could be degenerate. It is difficult to rule out either of the NBO models based on just the spectral modeling. 
Furthermore, the QPO rms spectrum and the covariance spectrum resembles a thermal spectrum. The covariance spectrum can be modeled using a single temperature blackbody with $kT\sim$ 2.8 keV and a multicolor blackbody model with $kT_{in}\sim$ 0.70 keV. The blackbody temperature obtained from the covariance spectrum agrees with that estimated from the energy spectrum.

If we consider the possibility of NBOs being the viscous oscillations of the spherical shell around the NS surface \citep{hasinger1987,titarchuk2001}, then we can constrain the size of this structure around the NS surface based on equation 1 from \citet{titarchuk2001} as given below:
\begin{equation}
L_{s} = \frac{f \nu_{s}}{\nu_{ssv}}
\end{equation}
Here, $\nu$$_{ssv}$ is the spherical shell viscous frequency (NBO frequency in this case), $f$ is 0.5 for the stiff and 1/2$\pi$ for the free boundary conditions in the transition layer, $\nu$$_{s}$ is the sonic velocity estimated based on the equation given below \citep{hasinger1987}:
\begin{equation}
\nu_{s}= 4.2 \times 10^{7} R_{6}^{-1/4} (\frac{M}{M_{\odot}} \frac{L}{L_{Edd}})^{1/8} cm \; s^{-1}
\end{equation}
where R$_6$ is the NS radius (units of 10$^6$ cm). 

For the $\sim$ 5.41 Hz NBO in the observation, based on the obtained flux and therefore luminosity from the spectral model, one can estimate the size of the oscillating shell to be $\sim$ 39 km ($f$=0.5). If the same oscillation causes an FBO then for a 20 Hz FBO, the shell size would be $\sim$ 10.5 km.
Further studies of the lower energy regime of NBOs are required to differentiate between the proposed models in a systematic manner.

\subsection{A comparison with previously studied NBOs}

\citet{wang2012} found in their investigation of Sco X-1 using RXTE/PCA data that rms amplitude of the NBO increases with photon energy below 13 keV, indicating that the strength of the NBO increases with energy and they note that the NBO strength reaches a plateau above 13 keV. This behavior is also seen in the NBOs detected in GX 340+0 \cite{jonker2000,pahari2024}. \citet{wang2012} also note a non-monotonic energy dependence of the NBO central frequency which they suggest could be due to a radial variation of the NBO emission region during the accretion process. 

In the case of Cyg X-2, \citet{mitsuda1989} found that the fractional rms increased with energy below 5--6 keV where it showed a pivot after which it increased with respect to the photon energy. This behavior is also seen in our study, where fractional rms and covariance drops significantly above 3 keV and there is a pivot above 5 keV. But above 5 keV the error bars are rather large as the effective area of NICER declines steeply at higher energies. Furthermore we find that the NBO feature is not seen in the 3--10 keV energy band, with an rms upper limit $<$ 1.1 \% at the QPO frequency in the NuSTAR data and $<$ 2.2 \% in the NICER 3--10 keV data with a 90 \% confidence level. Based on the different behavior of NBOs in soft and hard energy bands, we consider a possible scenario for the origin of these features to be associated with different variability components in the different energy bands.

The exploration of NBOs in the softer energy domain as is shown in this work has been limited up till now, but with NICER and NuSTAR we have an opportunity to explore the energy dependence of these features in the $<$ 3 keV energies simultaneously with the higher energy bands.\\
\\
\subsection{CCF lag studies}

CCF studies between the soft and hard energy bands of 0.5--3 keV vs 3--10 keV (NICER low and NuSTAR/NICER high energy bands) and 3--5 keV vs 3--10 keV (NuSTAR low vs NuSTAR high energy bands) were carried out using the `crosscor' tool in the {\it ftools} package. Previous studies \citep{sriram2019,malu2020,malu2021} suggest that CCF lags of a few hundred seconds between the soft and hard energy bands are possible readjustment timescales of the vertical Comptonization structure/corona. Our studies did not yield any significant CCF lags in the mentioned energy bands. This could mean that the compact corona is stable, without undergoing any drastic scale variations during the NB.

\section{Conclusion} \label{sec:conclusion}

We performed a spectro-temporal study of the NS LMXB Cyg X-2 using NICER and NuSTAR data. An energy dependent search for characteristic QPOs led to the detection of a NBO feature at $\sim$ 5.41 Hz. This feature appears in the middle portion of the NB branch and has a quality factor Q $\sim$ 2.57. We noted that the NBO appeared only in the 0.5--3 keV energy range but was absent in the 3--30 keV energy bands of NuSTAR and the 3--10 keV energy band of NICER data. Further binning of the 0.5--3 keV energy bands revealed that the feature peaks in the 1--2 keV energy band. NBO temperature, which is the temperature of the cool radial flow where the NBO originates, of Cyg X-2 has been previously identified to be $\sim$ 1 keV \citep{miller1992}. The energy spectrum of the source exhibits an excess emission feature at $\sim$ 1 keV, which has been previously reported by \cite{ludlam2022} and also seen in the other observations of Cyg X-2 \citep{vrtilek1986,vrtilek1988,kallman1989,smale1993,kuulkers1997}. This is considered to be the Fe L transition line originating from the photoionized plasma far away from the central source due to density fluctuations caused by the radiation pressure feedback of the central source.  Based on prior estimations of the 6 Hz NBO using a radiation hydrodynamic model, it is possible for the signature to be originating at a location $\sim$ 300 km from the central source. This, coupled with the consideration that the strength of the Fe L line depends on the density and temperature of the illuminated corona \citep{liedahl1990,kallman1995}, we suggest that the radius at which the Fe L emission and the NBOs originate may be within the same range and produced by a common underlying mechanism. Thus, lags seen in the frequency/energy dependent lag spectra of the source could be considered as the lags between soft and hard energy bands arising in the photoionized material far from the central source.
We study the frequency and energy dependent lag spectra of the source, which exhibited a few milliseconds hard lag in the frequency range that showed signatures of the NBO and a switch from hard to soft lags at $\sim$ 1 keV. The 12-15 millisecond hard lags seen in the NBO frequency range can not directly be interpreted as a variation occurring hundreds of km from the central source, but it could be indicative of variations occurring in the BL close to the source which causes variations in the radiation pressure feedback that results in the optical depth oscillations further out in the disk.
The rms and covariance spectrum peaks at 1--2 keV, further confirming previous findings. Search for similar soft energy NBO features in other NICER+NuSTAR observations of this source and other NS LMXBs are part of a project that is currently being conducted.

\section{Acknowledgements}
We thank the anonymous referee for their detailed feedback that has enhanced the scientific rigor of the manuscript. This research has made use of data and/or software provided by the High Energy Astrophysics Science Archive Research Center (HEASARC), which is a service of the Astrophysics Science Division at NASA/GSFC. This research has made use of data from the NuSTAR mission, a project led by the California Institute of Technology, managed by the Jet Propulsion Laboratory, and funded by the National Aeronautics and Space Administration. Data analysis was performed using the NuSTAR Data Analysis Software (NuSTARDAS), jointly developed by the ASI Science Data Center (SSDC, Italy) and the California Institute of Technology (USA). JH acknowledges support from NASA under award number 80GSFC21M0002.

\end{document}